\documentclass[%
 reprint,
 bibnotes,
 amsmath,amssymb,
 aps,
]{revtex4-2}

\usepackage{graphicx}% Include figure files
\usepackage{dcolumn}% Align table columns on decimal point
\usepackage{bm}% bold math
\usepackage{braket}
\usepackage{comment}
\usepackage{xcolor}
\usepackage[normalem]{ulem}

\begin{document}

\preprint{APS/123-QED}

\title{Spatial distribution of two symmetric four-wave mixing signals induced by Gaussian beams}

\author{M. R. L. da Motta}
\author{A. A. C. de Almeida}
\author{S. S. Vianna}
\email{sandra.vianna@ufpe.br}
\affiliation{Departamento de Física, Universidade Federal de Pernambuco, 50670-901, Recife, Pernambuco, Brazil\\}%

\date{\today}% It is always \today, today,

\begin{abstract}
We present a theoretical analysis of the spatial shape of two symmetric signals of degenerate four-wave mixing induced by Gaussian beams in a thin sample of two-level atoms.
Our calculations take into account the full spatial and spectral dependencies of the relevant nonlinear susceptibilities that govern the two processes.
This reveals two interesting effects.
The first one is that the total power of incident beams affects the transverse profile of the four-wave mixing signals at the medium exit and their free propagation.
The second one is the influence of the spectral characteristics of the medium on the longitudinal profile of both generated signals upon free propagation.
We argue that the first effect can be seen as the saturation of the medium in regions of higher intensity, while the second can be understood as the result of a nonlinear contribution to the refractive index inside the atomic medium.
These effects can be symmetric between the two signals, with asymmetries induced by different detunings from resonance of the incident fields.
\begin{description}
\item[Usage]
\item[Structure]
\end{description}
\end{abstract}

\maketitle

\section{Introduction}

The fundamental solution to the paraxial wave equation (PWE), the Gaussian beam, encompasses important properties of the actual output of laser sources, such as their characteristic intensity profile and phase distribution, both of which evolve as the beam propagates.
One exact higher-order solution to the PWE is the Hermite-Gaussian (HG) mode, which is given in cartesian coordinates and is characterized by lobes of light disposed in a rectangular grid. HG modes are important in the study of laser cavities, as they describe the spatial distribution of the output beams from these light sources \cite{kogelnik66}.
The Laguerre-Gaussian (LG) mode is another exact solution to the PWE. It is given in cylindrical coordinates and carries a ring-shaped intensity distribution.
Most importantly, LG modes carry well-defined orbital angular momentum (OAM) in the propagation direction per photon, as first demonstrated by Allen \textit{et al.} \cite{allen92}.
HG and LG modes can be represented in terms of one another and this correspondence is well-known \cite{allen92,kimel93}. Also, they are limiting cases of the so-called Ince-Gaussian paraxial mode \cite{bandres04}, that carries elliptical symmetry.
There are many other higher-order solutions to the PWE that present vastly diverse characteristics and interesting properties \cite{forbes2021}.

The above discussion considers the radiation field of light, in its many accessible spatial distributions or modes, propagating in free space.
Here, we are interested in the nonlinear light-matter interactions taking place in atomic media and in understanding how these interactions affect the spatial properties of the light field.
We will be concerned with a particular configuration of four-wave mixing (FWM), a third-order nonlinear optical process that can take place in a variety of systems, such as atomic vapors, cold atom samples, optical fibers, and others.
FWM has allowed the investigation of several optical phenomena, for instance, AC Stark shift, phase conjugation and electromagnetically induced transparency \cite{harter1980nearly,boyd81,yariv1977amplified,abrams1978degenerate,fleischhauer05}.
It has also been widely employed to study the interaction of structured optical modes with matter.

One setting that is commonly used to explore the spatial degrees of freedom of light is that of FWM induced by amplified spontaneous emission in a hot atomic vapor, with a 3-level cascade system \cite{chopinaud2018high,walker12,akulshin2015distinguishing,offer2018spiral,offer2020gouy}.
In cold atomic samples, FWM was employed to transfer OAM from incident to generated beams in nondegenerate \cite{tabosa1999optical} and degenerate \cite{barreiro2003} atomic systems, to transfer more complicated phase structures (obtained by superimposing LG modes of different orders) \cite{barreiro2004four}, and to store the information carried by the spatial structure of light in the ensemble of atoms, and later retrieve it \cite{moretti2009collapses,ding13}.

When the primary objective is the study of the spatial shape of the FWM beam, the usual approach is based on the overlap integral of four paraxial modes \cite{walker12,lanning17,offer2018spiral,offer2020gouy}.
In these cases, the calculations are performed regarding the interaction medium as a channel for the nonlinear process to take place, not possessing degrees of freedom that can affect the output mode superposition. In other words, the spatial distribution of the FWM field is fully determined solely by the distributions of the participating beams.
With this approach, the theoretical predictions are remarkably accurate \cite{walker12,offer2020gouy,lanning17}.
The role of the spatially dependent nonlinear coherence in the signal generation process was discussed in Ref. \cite{mallick20}.
In Ref. \cite{hamedi2018exchange}, the full spatial dependence of medium quantities is taken into account in calculations, and effects of detunings from resonance and phase mismatch on the phase distribution of the FWM beam are evidenced.
No connection is established, however, between the medium quantities and the spatial properties of the beam outside the interaction medium.
This is the point our work seeks to highlight.

This work is a theoretical study of the spatial shape of two symmetric signals of degenerate four-wave mixing generated in a sample of cold two-level atoms.
Our focus is on the influence of the spatially dependent nonlinear susceptibility, induced as a result of the nonlinear process, on the overall shape of the two FWM signals.
We investigate the spatial properties of the generated beams in various configurations, including different detunings from the atomic resonances and several intensities.
In particular, we show that the total power of incident beams affects not only the transverse profile of both FWM fields at the medium exit, but also the evolution of this profile under free propagation.
We show that the so-called root mean square (rms) parameters of the generated beam, which serve as an effective measure of the longitudinal profile, are intuitively affected by the frequency degrees of freedom of the atomic medium, in a way that resembles the Kerr effect.
These effects indicate that even in a thin medium, the power and frequency of pump beams may lead to significant modifications on the transverse and longitudinal profiles of a nonlinear signal.

Our results focus on the thin-medium regime, characterized by a medium length much smaller than the Rayleigh ranges of the participating fields. This regime is in agreement with the experimental conditions in a cold atom sample obtained with a magneto-optical trap (MOT).
On the other hand, in the thick-medium regime, the requirement of the Gouy phase-matching condition \cite{offer2020gouy} can lead to different results.
This regime is more easily achievable in a hot atomic vapor cell, for example.

This paper is divided as follows. In Sec. \ref{model}, we present details on the calculations performed, more specifically the semi-classical modeling of the nonlinear interaction, the solution to the wave equation for the FWM beam, and the properties of this solution.
In Sec. \ref{results} we present and discuss the main results of this work.
Finally, we establish our concluding remarks in Sec. \ref{conclusions}.

\section{Four wave mixing in a two-level system} \label{model}

The theoretical model used to describe the generated FWM beam can be divided in two main parts. The first one is the semi-classical description of the interaction between the atomic medium and the radiation field of the laser beams via optical Bloch's equations.
In particular, we are interested in the case where both incident beams are strong and possess the same power.
The second part consists on solving the nonhomogeneous wave equation for the FWM field $\mathcal{E}_s$ with the source term given by the nonlinear polarization vector $\mathbf{P}_{\mathrm{NL}}$, related to the nonlinear coherence obtained in the first part.

\subsection{Nonlinear light-atom interaction}

\begin{figure}[b]
\includegraphics[width=0.8\linewidth]{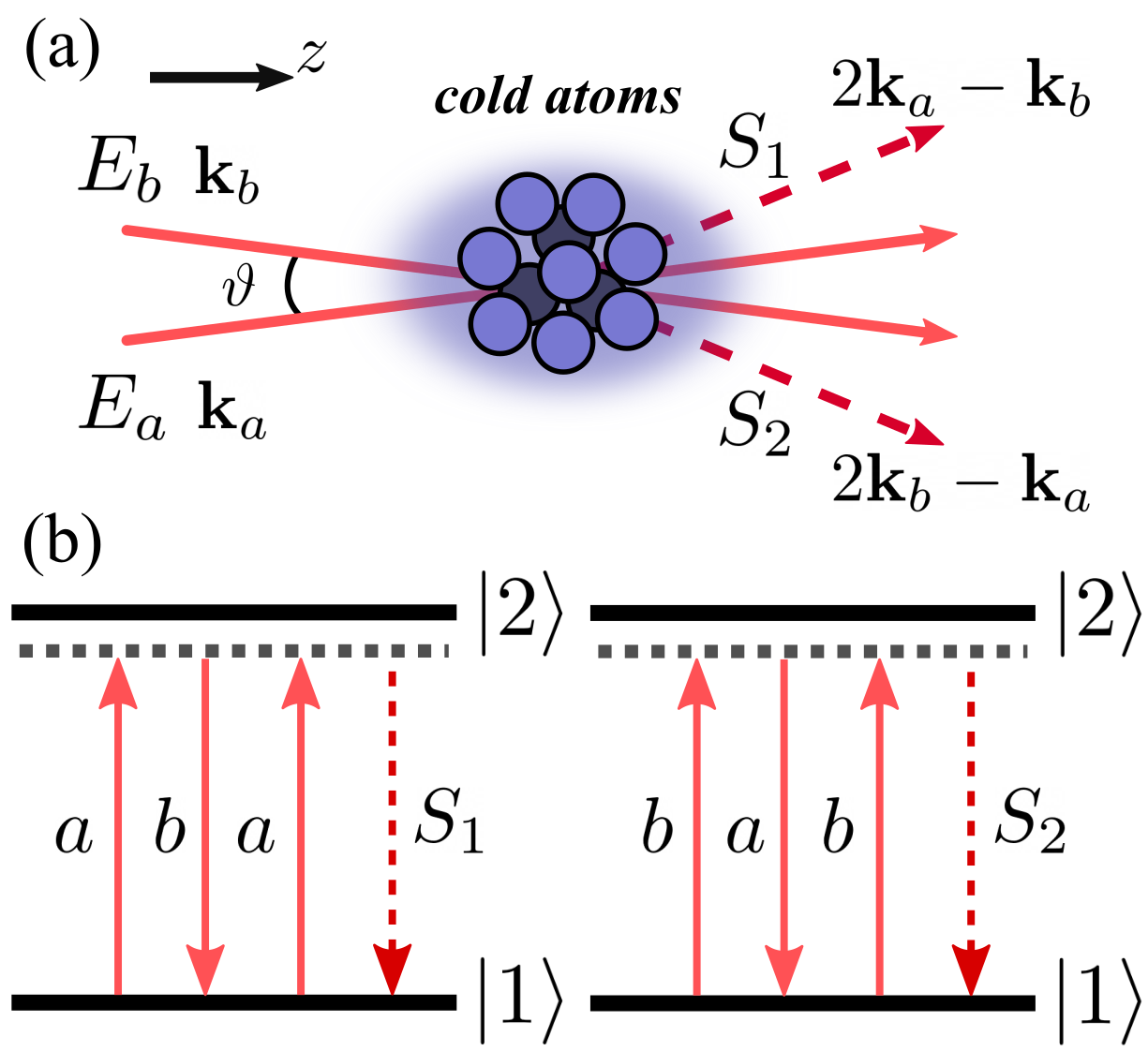}
\label{fig:lvl_diagram_subim_2}
\caption{(a) Spatial orientation of incident and FWM fields. (b) Depiction of the parametric processes that generate signals $S_1$ and $S_2$ in a two-level atom.}
\label{fig:beams_2lvl}
\end{figure}

We consider that both nonlinear signals are induced by two almost co-propagating Gaussian laser beams, with wave-vectors $\mathbf{k}_a$ and $\mathbf{k}_b$, and detected in the $2\mathbf{k}_a-\mathbf{k}_b$ and $2\mathbf{k}_b-\mathbf{k}_a$ directions, as shown in Fig. \ref{fig:beams_2lvl}(a).
We employ the density operator formalism to calculate the atomic medium response related to the nonlinear processes represented in Fig. \ref{fig:beams_2lvl}(b). We follow closely the development of Ref. \cite{boyd81}.
The total Hamiltonian is $\hat{H}=\hat{H}_{o}+\hat{H}_{\mathrm{int}}$, where $\hat{H}_{o}$ is the free-atom Hamiltonian, such that for the ground $\ket{1}$ and excited $\Ket{2}$ states, we have $\hat{H}_{o}\Ket{j}=\mathbb{E}_{j}\Ket{j}$, where $\mathbb{E}_{j}$ is the energy eigenvalue of $\ket{j}$, $j\in\{1,2\}$.
The interaction Hamiltonian is $\hat{H}_{\mathrm{int}}=-\hat{\boldsymbol{\mu}}\cdot\mathbf{E}\left(\mathbf{r},t\right)$, where $\hat{\boldsymbol{\mu}}=e\hat{\mathbf{r}}$ is the electric dipole operator and the total electric field is given by $\mathbf{E}(\mathbf{r},t)=\sum_{\upsilon} \mathbf{E}_{\upsilon}(\mathbf{r},t)$, $\upsilon\in\{a,b\}$,
\begin{equation}\begin{split} \label{eq:Efield}
    \mathbf{E}_{\upsilon}\left(\mathbf{r},t\right)&=\frac{1}{2}\boldsymbol{\epsilon}_{\upsilon}\mathcal{E}_{\upsilon}\left(\mathbf{r}\right)e^{-i\left(\mathbf{k}_{\upsilon}\cdot\mathbf{r}-\omega_{\upsilon}t\right)}+c.c.,
    \\
    &=\boldsymbol{\epsilon}_{\upsilon} E_{\upsilon}(\mathbf{r},t) + c.c.,
\end{split}\end{equation}
where $\mathbf{k}_\upsilon$ is the wave vector, $\boldsymbol{\epsilon}_{\upsilon}$ is the polarization direction, the amplitudes $\mathcal{E}_{\upsilon}(\mathbf{r})$ carry the transverse dependence of the fields and $c.c.$ means the complex conjugate. We consider a quasi co-propagating configuration, where the angle $\vartheta$ between $\mathbf{k}_a$ and $\mathbf{k}_b$ (Fig. \ref{fig:beams_2lvl}(a)) is very small, making $\mathbf{k}_{\upsilon}\cdot \mathbf{r} \simeq k_{\upsilon} z$, where  $k_{\upsilon}=|\mathbf{k}_{\upsilon}|=\omega_{\upsilon}n_\upsilon/c$ and $n_\upsilon$ is the index of refraction at frequency $\omega_\upsilon$.

The polarization directions $\boldsymbol{\epsilon}_a$ and $\boldsymbol{\epsilon}_b$ of input beams $\mathbf{E}_a$ and $\mathbf{E}_b$, respectively, determine the number of atomic states involved in the nonlinear process. We are interested in the case where $\boldsymbol{\epsilon}_a$ and $\boldsymbol{\epsilon}_b$ are parallel circular polarizations. Thus, the light-atom interaction can be described in terms of a two-level system.
We consider the $\Ket{5S_{1/2},F=2}\rightarrow\Ket{5P_{3/2},F=3}$ hyperfine transition of $^{87}\mathrm{Rb}$.
The nonlinear interaction leads to the generation of signal $S_1$, due to the absorption of two photons from beam $E_a$ and the stimulated emission of one photon from beam $E_b$; and signal $S_2$, due to the absorption of two photons from $E_b$ and the stimulated emission of one photon from $E_a$.
Figure \ref{fig:beams_2lvl}(b) illustrates these processes schematically.

We define the Rabi frequency
\begin{equation}
\begin{split}
    \Omega_{\upsilon}(\mathbf{r})=\dfrac{\mu_{jk}\mathcal{E}_{\upsilon}\left(\mathbf{r}\right)}{2\hbar},
\end{split}
\end{equation}
where $\mu_{jk}=\bra{j}(\hat{\boldsymbol{\mu}}\cdot\boldsymbol{\epsilon}_{\upsilon})\ket{k}$, with $j\neq k$, is the electric dipole matrix element (due to parity, $\mu_{jj}=0$). The matrix elements of $\hat{H}_{\mathrm{int}}$ can be writtten
\begin{equation} \label{eq:H_int}
    H_{\mathrm{int},jk}=-\hbar\sum_{\upsilon}\Omega_{\upsilon}e^{-i(k_{\upsilon}z - \omega_{\upsilon}t)} + c.c.
\end{equation}
The density operator $\hat{\rho}=\sum_{jk}\rho_{jk}\ket{j}\bra{k}$ describes the state of the atomic ensemble and satisfies $\sum_{j}\rho_{jj} = 1$, and $\rho_{jk} = \rho_{kj}^*$, where $({}^*)$ means complex conjugation. Its time evolution is given by Liouville's equation with a relaxation term $\hat{\mathcal{L}}$ associated with spontaneous decay from $\ket{2}$ to $\ket{1}$
\begin{equation}
    \frac{d \hat{\rho}}{d t}=\dfrac{i}{\hbar}[\hat{\rho},\hat{H}] + \hat{\mathcal{L}}.
\end{equation}
In our model, populations and coherences decay at rates $\Gamma$ and $\Gamma/2$, respectively, where $\Gamma/2\pi$ is the natural decay rate of the excited state.
For the $D_2$ line of $^{87}$Rb, $\Gamma/2\pi \approx 6$ MHz \cite{steck2001rubidium}.
With this, we obtain the optical Bloch's equations (OBEs) for the two-level system
\begin{equation} \label{eq:OBEs}
    \begin{split}
        \dot{(\Delta\rho)} = &-\dfrac{2i}{\hbar}[\rho_{12}H_{\mathrm{int},21}-c.c.] -\Gamma\left[\Delta\rho-(\Delta\rho)^{0}\right],
        \\
        \dot{\rho}_{12}	= &-\dfrac{i}{\hbar}\left[H_{\mathrm{int},12}\Delta\rho-\rho_{12}\left(\mathbb{E}_{2}-\mathbb{E}_{1}\right)\right] - \dfrac{\Gamma}{2}\rho_{12},
    \end{split}
\end{equation}
where $\Delta\rho=(\rho_{22}-\rho_{11})$ is the population difference and $(\Delta\rho)^0$ is the population difference far from the region of interaction with fields $E_a$ and $E_b$.
We assume that the coherence $\rho_{12}$ oscillates with frequencies $\omega_a$, $\omega_b$ and $2\omega_a-\omega_b$ \cite{boyd81,harter1980nearly}
\begin{equation} \label{eq:coh_2lvl}
\begin{split}
    \rho_{12} &= \sigma_{12}^a e^{i\omega_at}+ \sigma_{12}^be^{i\omega_bt} +\sigma_{12}^{2a-b}e^{i
    (2\omega_a-\omega_b)t} ,
\end{split}
\end{equation}
where $\sigma_{ij}$ are the slowly varying coherences.
The $2\omega_a - \omega_b$ component is responsible for the FWM process that generates the signal $S_1$.
The population difference $\Delta\rho$ has a stationary component and one oscillating at $|\omega_a-\omega_b|$ \cite{harter1980nearly},
\begin{equation} \label{eq:pd_2lvl}
    \Delta\rho=(\Delta\rho)^{\mathrm{dc}} + \left[(\Delta\rho)^{a-b}e^{i(\omega_a-\omega_b)t} +c.c.\right].
\end{equation}
We now substitute equations (\ref{eq:H_int}), (\ref{eq:coh_2lvl}) and (\ref{eq:pd_2lvl}) into Eqs. (\ref{eq:OBEs}), perform the rotating wave approximation and collect terms that oscillate with the same frequency. Then, in the steady state regime we arrive at the set of algebraic equations shown in Appendix \ref{appendix:RHO} for the slowly varying coherences and population differences.
Defining $\widetilde{\Omega}_{\upsilon} \equiv \Omega_{\upsilon} e^{-i k_{\upsilon} z}$, $\delta_{\upsilon} \equiv \omega_{\upsilon}-\omega_o$ as the detuning from resonance of field $\Omega_{\upsilon}$ and $\omega_o \equiv (\mathbb{E}_2-\mathbb{E}_1)/\hbar$ as the resonance frequency, we obtain
\begin{equation} \label{eq:s12}
    \sigma_{12}^{2a-b}=\dfrac{-2i\widetilde{\Omega}_{a}^{2}\widetilde{\Omega}_{b}^{*}(\Delta\rho)^{\mathrm{dc}}(1/\Delta_a+1/\Delta_b)}{(\Delta_a+\Delta_b)\left(2i\delta_{a}-i\delta_{b}+\Gamma/2\right)+2|\Omega_{a}|^{2}},
\end{equation}
    where $\Delta_a = i\delta_{a}+\Gamma/2$, $\Delta_b = -i\delta_{b}+\Gamma/2$ and
\begin{equation}
    (\Delta\rho)^{\mathrm{dc}} = \dfrac{(\Delta\rho)^0}{1 + \dfrac{2|\Omega_a|^2}{\delta_a^2+\Gamma^2/4} + \dfrac{2|\Omega_b|^2}{\delta_b^2+\Gamma^2/4}}.
\end{equation}
The coherences $\sigma_{12}^a$ and $\sigma_{12}^b$ can be found as
\begin{align}
    \sigma_{12}^a&=\dfrac{-i\widetilde{\Omega}_a\rho_{11}^0}{\Delta_a+2|\Omega_a|^2/\Gamma+2|\Omega_b|^2/\Delta_{ab}}, \label{eq:sigma_a}
    \\
    \sigma_{12}^b&=\dfrac{-i\widetilde{\Omega}_b\rho_{11}^0}{\Delta^*_b+2|\Omega_b|^2/\Gamma+2|\Omega_a|^2/\Delta_{ab}^*}, \label{eq:sigma_b}
\end{align}
where $\Delta_{ab}=i\delta_a-i\delta_b+\Gamma/2$ and $\rho_{11}^0=1$ is the population of the ground state far from the region of interaction.

We rewrite the coherences given by Eqs. (\ref{eq:s12}), (\ref{eq:sigma_a}) and (\ref{eq:sigma_b}) in the forms
\begin{align}
        \sigma_{12}^{2a-b} &= X^{2a-b} \widetilde{\Omega}_a^2\widetilde{\Omega}_b^*,
        \\
        \sigma_{12}^a &= X^a \widetilde{\Omega}_a,
        \\
        \sigma_{12}^b &= X^b \widetilde{\Omega}_b,
\end{align}
where $X^{2a-b}$, $X^a$ and $X^b$ are the couplings associated with the processes in directions $(2\mathbf{k}_a-\mathbf{k}_b)$, $\mathbf{k}_a$ and $\mathbf{k}_b$, respectively.
These coupling factors carry all the spectral response of the medium.
They also depend on the field amplitudes $|\Omega_{a,b}|^2$, and thus the corresponding susceptibilities contain information regarding higher order processes in the same direction.
We define the effective susceptibilities
\begin{align}
    \chi^{2a-b} &= \dfrac{\mathcal{N}|\mu_{12}|^4}{\varepsilon_o\hbar^3} X^{2a-b}, \label{eq:chi_2a_b}
    \\
    \chi^{ab} &=\dfrac{\mathcal{N}|\mu_{12}|^2}{\varepsilon_o\hbar} (X^a+X^b), \label{eq:chi_ab}
\end{align}
where $\mathcal{N}$ is the atomic density.
Since the atoms are considered stationary, we do not need to include the effect of Doppler broadening.

\subsection{Wave equation}

We are interested in the field distributions of the generated signals, and thus seek a solution to the wave equation for field $\mathbf{E}_{s}$. We focus only on $S_1$, since the equations for $S_2$ are obtained and solved in the same way. The wave equation for the FWM electric field is \cite{jackson1999classical}
\begin{equation} \label{eq:wave_eq_start}
    \nabla^{2}\mathbf{E}_{s}-\dfrac{1}{c^{2}}\dfrac{\partial^{2}\mathbf{E}_{s}}{\partial t^{2}}=\mu_{o}\dfrac{\partial^{2}\mathbf{P}}{\partial t^{2}},
\end{equation}
where $\mathbf{E}_s$ is written as in Eq. (\ref{eq:Efield}).
The macroscopic polarization $\mathbf{P}$ can be divided in two components: $\mathbf{P}_{2a-b}$, which describes the generation process of the FWM field $E_s$; and $\mathbf{P}_{ab}$, associated with the propagation of the generated field inside the medium affected by fields $E_a$ and $E_b$. We can write the total polarization as $\mathbf{P}=\mathbf{P}_{ab} + \mathbf{P}_{2a-b}$, and its projection onto the oscillation direction of the generated field is
\begin{equation} \label{eq:Pol}
    (\mathbf{P}\cdot\boldsymbol{\epsilon}^*_s) = \varepsilon_o\chi^{ab}E_s + \varepsilon_o\chi^{2a-b}E_a^2E_b^*.
\end{equation}
In the calculations that follow, and throughout the rest of this work, we neglect the first term on the right-hand side of Eq. (\ref{eq:Pol}). This is done under the thin-medium regime assumption, such that the generated field is not influenced by the medium response associated with $\chi^{ab}$.
Under the rotating wave and paraxial approximations, we obtain
\begin{equation} \label{eq:wave_eq3}
    \left(\frac{i}{2k_{s}}\nabla^{2}_{\perp}+\frac{\partial}{\partial z}\right)\mathcal{E}_s  = \kappa\Omega_a^2\Omega_b^* e^{-i\Delta k z},
\end{equation}
where $\nabla^2_\perp$ is the transverse Laplacian, $\Delta k = |2\mathbf{k}_a - \mathbf{k}_b - \mathbf{k}_s| \simeq 2k_a - (k_b + k_s)\cos\vartheta$ is the phase mismatch and $\kappa$ is the nonlinear coupling, given by
\begin{align}\label{eq:kappa}
    \kappa(r,z;\delta)=-i\dfrac{\omega_s\hbar^3}{2c\mu_{12}|\mu_{12}|^2}\chi^{2a-b}(r,z;\delta).
\end{align}
In the last equation, $\delta$ represents $\delta_a$ and $\delta_b$.

We highlight that there is an implicit position dependence in $\kappa$.
It exists only inside the interaction region, between $z=-L/2$ and $z=L/2$, and is zero everywhere else in the $z$-axis.
Thus, Eq. (\ref{eq:wave_eq3}) describes the nonlinear signal generation process inside the sample.
There is no nonlinear signal at positions $z\leq -L/2$, and so the boundary condition is $\mathcal{E}_s(\mathbf{r}_\perp,-L/2)=0$.
For $z>L/2$, where the generated beam propagates in free space, $\mathcal{E}_s$ must satisfy the homogeneous paraxial wave equation, $(i\nabla^2_\perp/2k_s+\partial/\partial z)\mathcal{E}_s=0$, with the boundary condition given by the solution of Eq. (\ref{eq:wave_eq3}) at $z=L/2$, $\mathcal{E}_s(\mathbf{r}_\perp,L/2)$.

For the solution of Eq. (\ref{eq:wave_eq3}), we first note that the incident beams are strong, and we expect them to undergo little extinction and transverse structure variation along the interaction region.
Thus, we treat Eq. (\ref{eq:wave_eq3}) uncoupled from the wave equations for $\mathcal{E}_a$ and $\mathcal{E}_b$. This simplifies the mathematical work of our problem and is shown to bring results with good agreement with experimental measurements \cite{lanning17,offer2020gouy,walker12}.

\subsection{Solution to the FWM field wave equation}

It is well known that both the HG and LG paraxial modes form complete orthonormal sets of functions on the transverse plane.
This property allows to write any scalar optical field as a superposition of the form $U(\mathbf{r}_\perp,z)=\sum_{m,n}c_{mn}(z)u_{mn}(\mathbf{r}_\perp,z)$, where indices $(m,n)$ characterize the modes of the chosen basis and, in analogy with a quantum-mechanical system, the expansion coefficients $c_{mn}(z)$ can be seen as probability amplitudes of finding the system $\Ket{U}$ in the state $\ket{m,n}$.

We consider only incident Gaussian beams, and therefore, the LG basis is most convenient due to its cylindrical symmetry.
The LG mode is denoted as
\begin{equation} \label{eq:uLG_2}
\begin{split}
    u_{\ell p}(r,\phi,z) &= \frac{C_{\ell p}}{w(z)}\left(\frac{\sqrt{2}r}{w(z)}\right)^{|\ell|} L_{p}^{|\ell|}\left[\frac{2r^2}{w^2(z)}\right] e^{i\ell\phi}
    \\
    &  \times  e^{-\frac{r^2}{w^2(z)}} \exp\left[- i\frac{k r^{2}}{2R(z)} + i\Psi_{\mathrm{G}} (z) \right],
\end{split}
\end{equation}
where $C_{\ell p}=\sqrt{2p!/\pi(p+|\ell|)!}$ is the normalization constant, $L_{p}^{|\ell|}(\cdot)$ is the associated Laguerre polynomial, $w(z)=w_{o}\sqrt{1+(z/z_{R})^{2}}$ is the beam waist, $R(z)=z\left[1+(z_{R}/z)^{2}\right]$ is the curvature radius, $\Psi_{\mathrm{G}}(z)=(N_{\ell p}+1)\tan^{-1}(z/z_R)$ is the Gouy phase shift, with the total mode order defined as $N_{\ell p}=2p+|\ell|$, $z_{R}=kw_{o}^{2}/2$ is the Rayleigh range and $w_o$ is the minimum beam waist.

A light beam described by an LG mode carries well-defined OAM in the $z$-direction, which is related to the azimuthal phase factor $e^{i\ell \phi}$, where integer $\ell\in(-\infty,\infty)$, called the topological charge, defines the OAM per photon in the beam \cite{allen92}. The other index characterizing the mode, $p\in[0,\infty)$, is called the radial index.
It is related to the number of dark rings in the intensity profile of $u_{\ell p}$, but does not have a straightforward connection with a physical quantity as is the case for $\ell$.
In recent years, however, the radial index has been the subject of theoretical works \cite{karimi2014radial,plick2015} that have enlightened its significance.

We write the generated field amplitude $\mathcal{E}_s$ as the superposition
\begin{equation} \label{eq:field_exp}
\mathcal{E}_s (\mathbf{r}) = \sum_{\ell=-\infty}^\infty \sum_{p=0}^\infty \mathcal{A}_{\ell p}(z) u_{\ell p}(\mathbf{r}).
\end{equation}
The problem becomes that of finding the set of relevant coefficients $\{\mathcal{A}_{\ell p}\}$. Substituting Eq. (\ref{eq:field_exp}) into Eq. (\ref{eq:wave_eq3}) and employing the orthogonality relation of $u_{\ell p}$, we obtain an equation for $\mathcal{A}_{\ell p}(z)$.
\begin{equation} \label{eq:dAlpdz}
    \dfrac{\partial \mathcal{A}_{\ell p} (z)}{\partial z} = \Lambda_p^{\ell}(z)e^{-i\Delta kz},
\end{equation}
where
\begin{equation}\label{eq:Lambda}
    \Lambda_{p}^{\ell}(z) = \int_0^{2\pi}\int_0^{\infty} \kappa(\mathbf{r})\Omega_a^2\Omega_b^*u_{\ell p}^*rdrd\phi,
\end{equation}
is the projection of the spatially dependent nonlinear source term onto the LG function space, called the transverse overlap integral.

As we consider only Gaussian incident beams, that carry no OAM and thus possess $\ell = 0$, it will be useful to define the single index mode $u_p(r,z) \equiv u_{0p}(r,z)$, which is azimuthally symmetric.
Since $\kappa(\mathbf{r})$ only contains the squared modulus of fields $\Omega_{a,b}$, it has no $\phi$ dependence, even for $\ell_{a,b}\neq0$.
The azimuthal integral in Eq. (\ref{eq:Lambda}) yields $\int^{2\pi}_0 e^{i\ell\phi} d\phi = 2\pi\delta_{\ell,0}$, and $\Lambda^\ell_p(z)$ becomes $\Lambda_p(z)\equiv\Lambda^{0}_p(z)=2\pi\int^\infty_0\kappa\Omega^2_a\Omega^*_b u^*_p r dr$.
We then see that, as the incident fields have $\ell=0$, the FWM field does not carry OAM content, as anticipated, and can be written as a superposition of radial modes $\mathcal{E}_s=\sum_p \mathcal{A}_p(z)u_p(r,z)$, where $\mathcal{A}_p\equiv\mathcal{A}_{\ell=0,p}$.
The presence of modes with different orders $N_{p}=N_{0p}=2p$ can lead to the interference of the various Gouy phase factors $\exp{[i(1+N_{p})\tan^{-1}(z/z_R)]}$ that affects the radial structure of $|\mathcal{E}_s|^2$ upon propagation \cite{pereira17,wu20}.
When incident fields carry single topological charges or an arbitrary superposition of topological charges, the azimuthal integral imposes the conservation of OAM \cite{walker12,offer2018spiral,offer2020gouy}.
This is the subject of future work currently underway.

We then solve Eq. (\ref{eq:dAlpdz}) to find the expansion coefficients as
\begin{equation} \label{eq:Alp_z}
    \mathcal{A}_{p}(z) = \int_{-L/2}^{z_<} \Lambda_{p}(z') e^{-i\Delta k z'}dz',
\end{equation}
where $z_< = \min[z,L/2]$.
This solution is suitable for both regions of space: $-L/2<z\leq L/2$ and $z>L/2$.
We are interested in the FWM beam outside the medium, where it can be detected, and thus seek to evaluate $\mathcal{A}_{p}(L/2)$.
It is important to note that, although not explicit, $\mathcal{A}_{p}(z)$ is also a function of the detunings $\delta_a$ and $\delta_b$.

The characteristic length of $\Lambda_{p}(z)$ is given by the Rayleigh range of the beams that participate in the FWM process, $z_R$.
For a thin-medium, characterized by $L\ll z_R$, we can neglect the variation of $\Lambda_{p}(z)$ inside the interaction region and take its value at $z=0$.
This allows to remove the transverse overlap from the $z$ integral in Eq. (\ref{eq:Alp_z}) and write the approximate form
\begin{equation}
\begin{split}
    \mathcal{A}_{p}(L/2) &\simeq \Lambda_{p}(0) \int_{-L/2}^{L/2}  e^{- i\Delta k z'}dz',
    \\
    &= \Lambda_{p}(0) T(L).
\end{split}
\end{equation}
In the above expression, $T(L)=2\sin(\Delta k L/2)/\Delta k$ can be regarded as an efficiency measure of the signal generation process inside the medium.
It takes into account the phase-mismatch and is a common factor to all $\mathcal{A}_{p}$.
Thus, in the thin-medium regime, all the information of the nonlinear wave mixing process is contained in the transverse overlap integral evaluated at $z=0$, $\Lambda_p(0)$. It is therefore the main quantity in our calculations, and completely determines the distribution of modes of the generated fields.

The orthogonality of LG modes allows to write the total power of field $\mathcal{E}_s$ as $P = \sum_{p} |\mathcal{A}_{p}|^2$, which gives the normalization factor of $\mathcal{E}_s$ at fixed $\delta$.
We define the mode purity or mode weight as $\eta_p \equiv |\mathcal{A}_p|^2/P$, a measure of the relative contribution of the mode $u_{p}$ to $\mathcal{E}_s$.
The phase angle of $\mathcal{A}_{p}$ is $\Phi_{p}$, such that we can write the expansion coefficient in the normalized form $\mathcal{A}_{p} = \sqrt{P\eta_{p}}\,e^{i\Phi_{p}}$.
Note that the relative phases between the various expansion coefficients can also be responsible for changes on the output beam superposition.

The nonlinear coupling $\kappa(r,z;\delta)\propto\chi^{2a-b}(r,z;\delta)$ has a complicated dependence on the input fields, and, rigorously, on the position $\mathbf{r}$. 
However, even though this coupling can promote sensible modifications to the FWM beam in the near-field, it does not contribute substantially to the transverse shape of the \emph{far-field} FWM beam, which is mainly determined by the mode components of fields $\Omega_a$ and $\Omega_b$.
This suggests that the FWM beam profile is dictated by the overlap of input beams.
Indeed, this is usually assumed in the description of nonlinear processes involving beams with OAM or arbitrary transverse structures.
In references \cite{schwob1998transverse,pereira17,alves2018conditions,buono2020chiral}, that focus on second-order nonlinearities (parametric oscillation and second-harmonic generation), and in references \cite{walker12,offer2018spiral,offer2020gouy}, that treat FWM, the quantities analogous to that of Eq. (\ref{eq:Lambda}) are overlap integrals of three ($\chi^{(2)}$) and four ($\chi^{(3)}$) LG modes.
In all of these cases, the relevant nonlinear susceptibility, $\chi^{(2)}$ or $\chi^{(3)}$, is a uniform quantity that factors out of the overlap integral and does not affect the value of the mode expansion coefficients.
With the present study, we seek to understand the influence of the atomic medium on the FWM process.
More specifically, the role of the full spatial and spectral dependencies of $\chi^{2a-b}(\mathbf{r};\delta)$ on the spatial features of signals $S_1$ and $S_2$ in the particular FWM configuration considered.
Our results, in Sec. \ref{results}, show that by varying the power and the frequency of input fields, one can produce changes on the distributions of mode weights $\eta_{p}$ and phases $\Phi_{p}$, which in turn generate modifications on the FWM beams upon free space propagation.
This is because in our case, the effective susceptibilty can not be factored out from the overlap integral, and is taken fully into account in the calculations.

\subsection{Longitudinal profile of the FWM field}

The spot size of the FWM beam is well described by the rms radius, defined as $r_{\mathrm{rms}}(z)\equiv[\frac{1}{P}\iint r^2 |\mathcal{E}_s|^2 r dr d\phi]^{\frac{1}{2}}$ \cite{phillips83}.
Substituting Eq. (\ref{eq:field_exp}), at $z>L/2$ we can obtain the form \cite{vallone16}
\begin{equation} \label{eq:rms_par}
r_{\mathrm{rms}}(z)=\sqrt{r^2_m+\theta^2_{\mathrm{rms}}(z-z_m)^2},
\end{equation}
where $r_m$ is the minimum spot size, $z_m$ is the position where it occurs and $\theta_\mathrm{rms}$ is the divergence angle. These three parameters are calculated from the set of coefficients $\{\mathcal{A}_p\}$ (see Appendix \ref{appendix:rms}) and completely determine the longitudinal profile of the beam.
We can also introduce the dimensionless quantity $\mathcal{M}^2 = k\theta_{\mathrm{rms}}r_m$, called the beam quality factor, which is proven to satisfy the bound $\mathcal{M}^2\geq 1+\langle|\ell|\rangle$ \cite{vallone16}, where $\langle{|\ell|}\rangle$ is the mean value of the topological charge magnitude in the field expansion. In our case, $\langle|\ell|\rangle=0$ and the quality factor must therefore satisfy $\mathcal{M}^2\geq1$.

We highlight that $\theta_{\mathrm{rms}}$, $r_m$, $z_m$, and $\mathcal{M}^2$, the so-called rms quantities, depend on the detunings $\delta_{a,b}$.
In order to highlight that this dependence arises in the theory due to the spatial distribution of $\chi^{2a-b}$, we consider the spatially uniform nonlinear coupling $\overline{\kappa}(\delta)\propto\overline{\chi}(\delta)$, an averaged measure of the nonlinear response over the interaction volume.
Then, $\overline{\kappa}(\delta)$ factors out of the integral given by Eq. (\ref{eq:Lambda}), which becomes the integral of a product of four LG modes \cite{walker12,offer2018spiral,offer2020gouy}.
It can be seen that in this case all of the dependence on the frequency degrees of freedom factor out from the coefficients $\mathcal{A}_p$ and the rms quantities do not vary with $\delta$.

\section{Results and Discussion} \label{results}

In this section, we present and discuss our main results.
Together, they comprise a theoretical investigation of various aspects of the FWM signal generation (induced by Gaussian incident beams) and the free space propagation after leaving the nonlinear medium.

Our calculations consist of evaluating the expansion coefficients $\mathcal{A}_p$ for the two symmetric signals $S_1$ and $S_2$.
With the set $\{\mathcal{A}_{p}\}$, all of the quantities we are interested in can be obtained, such as the intensity distribution $|\Omega_s|^2$ and its propagation outside the medium, the mode components $\eta_{p}$, and others.
We do this for several configurations.
Here, each configuration is defined by the amplitudes $\Omega_{a}^0$ and $\Omega_b^0$, related to the total power of the incident beams ($\Omega_{a,b}(\mathbf{r})=\Omega^0_{a,b}u_0(\mathbf{r})$); and the detunings from resonance $\delta_{a}$ and $\delta_b$.
In the calculations, all beams possess the same wavelength $\lambda$ and minimum waist $w_o$, giving the same Rayleigh range $z_R$.

We consider that the waist of all beams is $w_o=1$ mm near the interaction region.
For the wavelength $\lambda = 780$ nm, used to excite the $D_2$ line of $^{87}$Rb, the Rayleigh range is $z_R=\pi w_o^2/\lambda\approx4$ m.
A cold atom cloud usually obtained with a MOT has a size $L$ of a few millimeters.
The condition $L/z_R\ll1$ is satisfied, thus justifying our focus on the thin medium regime.

As will be seen, for purely Gaussian incident beams, the transverse shapes of the FWM beams are mainly Gaussian but can have significant contributions from modes with $p>0$.
We show how the transverse profiles of the generated FWM beams and their free space propagation are affected by the total power contained in the incident beams and by the detunings from resonance $\delta_a$ and $\delta_b$.

\subsection{FWM induced by Gaussian beams}

For Gaussian input beams, the two signals $S_1$ and $S_2$ carry null topological charges, $\ell_1=\ell_2=0$. Also, for equal amplitudes $\Omega_a^0=\Omega_b^0$ and detunings $\delta_a=\delta_b=\delta$, they possess completely symmetric transverse shapes and free space propagation characteristics.
This is because the nonlinear coherence, given by Eq. (\ref{eq:s12}), remains the same with an exchange of labels $a\leftrightarrow b$.

\subsubsection{Effect of pump intensity}

\begin{figure*}[t!]
    \includegraphics[width=0.95\linewidth]{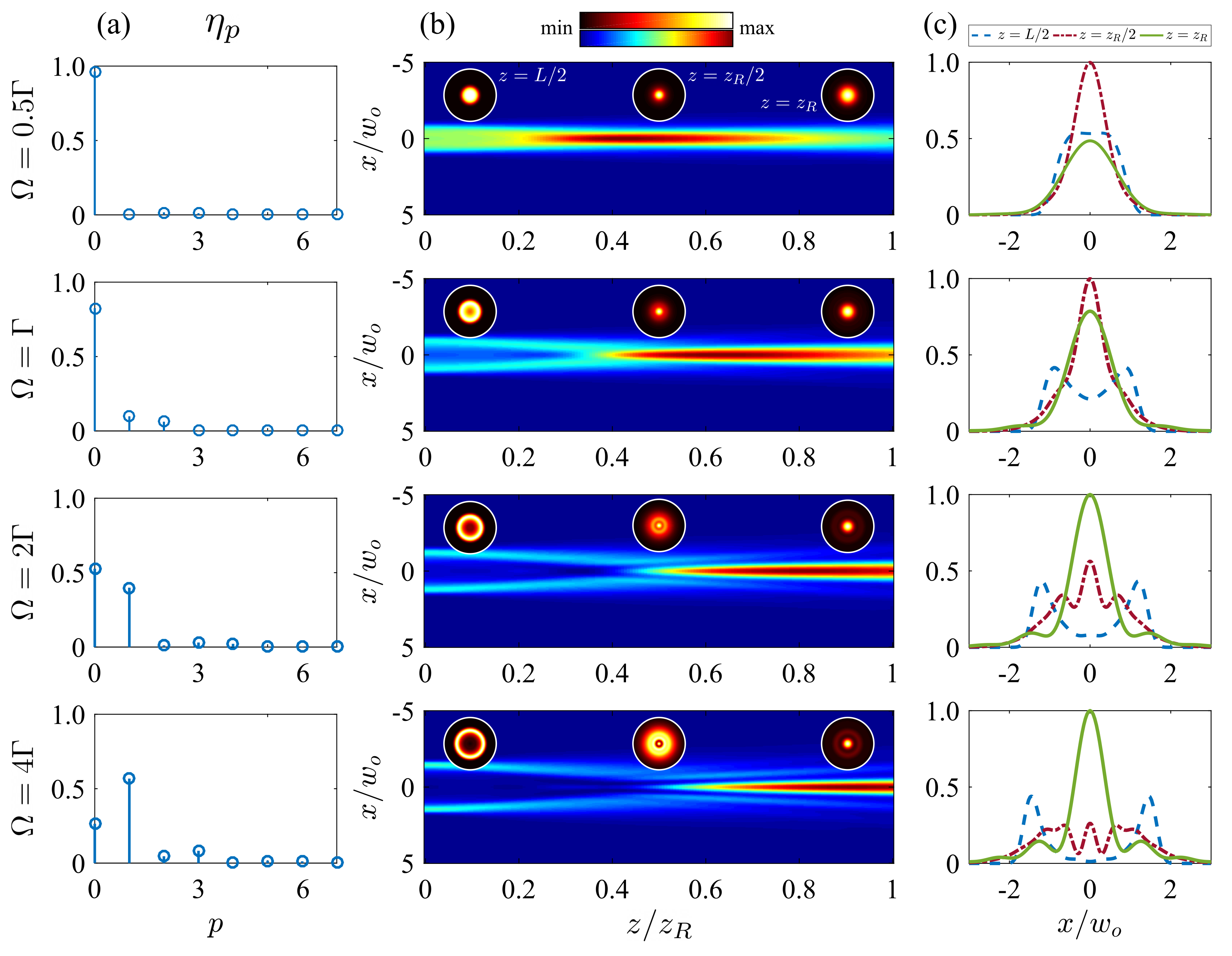}
    \caption{(a) Mode weights ${\eta}_{p}$, (b) propagation of the FWM beam intensity profile outside the interaction medium from $z/z_R=L/2z_R\approx0$ to $z/z_R=1$ (insets show the intensity profile at $z=L/2,z_R/2,z_R$) and (c) normalized radial distribution of intensity at positions $z=L/2, z_R/2$ and $z_R$, for incident Gaussian beams with Rabi frequencies $\Omega_{a,b}^0=\Omega=0.5\Gamma,\Gamma,2\Gamma$ and $4\Gamma$ (from top to bottom).}
    \label{fig:intensity_amplitudes}
\end{figure*}

We now investigate the effect of the total power contained in the incident beams on the overall shape of the generated beams.
The weights ${\eta}_{p}$, that quantify the contribution of each mode $u_{p}$ to the generated field $\mathcal{E}_s$, is represented in Fig. \ref{fig:intensity_amplitudes}(a), and the radial distribution of intensity at different longitudinal positions is discriminated in Fig. \ref{fig:intensity_amplitudes}(c).
Figure \ref{fig:intensity_amplitudes}(b) shows the FWM beam propagation for $\delta_a=\delta_b=0$ and different values of $\Omega_{a}^0 =\Omega_b^0=\Omega$. We note that as the amplitudes increase, the overall shape of the generated beam at the nonlinear medium exit changes significantly.
However, keeping $\Omega_a^0 = \Omega_b^0$, the symmetry between both generated beams is preserved.
Near the medium exit, $z/z_R=L/2z_R\approx0$, the beam becomes ring-shaped. We attribute this behavior to a spatial saturation effect, which can be understood by inspecting equation ($\ref{eq:s12}$). In all cases, the near-field intensity profile is determined by the nonlinear coherence, $|\sigma^{2a-b}_{12}|^2\propto |\chi^{2a-b}|^2I^2_aI_b$, and due to the Gaussian distribution of the fields, for greater amplitudes, the denominator in Eq. (\ref{eq:s12}) is larger at the center, making $|\sigma_{12}^{2a-b}|^2$ smaller in this region.
Nonetheless, after propagating distances of the order of $z_R$, the beams acquire a shape that corresponds to the dominant modes, as we can see in the radial profile at $z=z_R$ (green line) in Fig. \ref{fig:intensity_amplitudes}(c).
This is similar to what is verified in Ref. \cite{pereira17}.

For the lower intensities ($\Omega=0.5\Gamma,\Gamma$) the dominant mode is $u_0$, and the far-field profile is mainly Gaussian.
In this case, the saturation effects are very small, making the influence of the nonlinear coupling $\kappa(\mathbf{r})$ in Eq. (\ref{eq:Lambda}) negligible.
The values of $\eta_{p}$ approach those obtained by calculating the overlap integral of four LG modes \cite{walker12,offer2018spiral,offer2020gouy}.
On the other hand, for the higher intensities ($\Omega=2\Gamma,4\Gamma$) there is a significant contribution from the mode $u_1$, and a dark ring is present on the far-field profile.
This transition of the radial profile is due to the superposition of modes in the generated beam that possess different mode numbers $N_{p}$, and thus acquire different Gouy phases upon propagation \cite{pereira17,wu20}.

Sensible change of the longitudinal profile is achieved by increasing $\Omega$.
We highlight that the generated beam outside the medium clearly indicates an intensity maximum at positions $z/z_R$ far from $0$, for all incident beam amplitudes.
In fact, for increasing $\Omega$, this position of maximum intensity is shifted towards greater $z/z_R$, as seen from Fig. \ref{fig:intensity_amplitudes}(b).
This would make one expect the position of minimum $r_{\mathrm{rms}}(z)$ to be shifted as well.
However, this is not the case, since the minimum of $r_{\mathrm{rms}}(z)$ does not necessarily correspond to the position of maximum local intensity.
This is because $r_\mathrm{rms}(z)$ for an arbitrary beam does not correspond to the radial position where the field amplitude decreases by a factor of $1/e$ relative to the amplitude at the center, as it does for a pure Gaussian beam.

\subsubsection{Effect of detunings from resonance}

\begin{figure}[b!]
\centering
\includegraphics[width=1\linewidth]{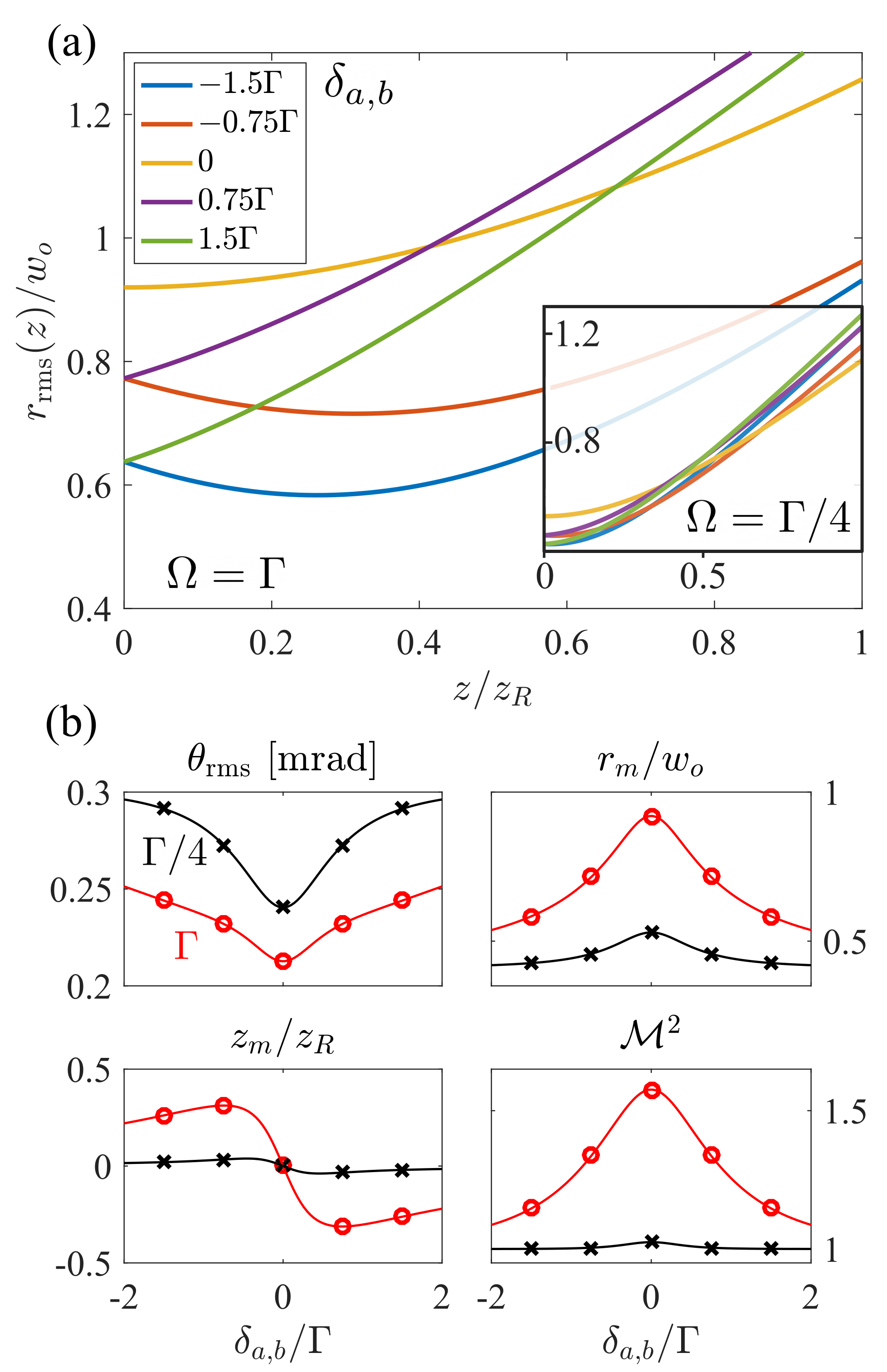}
\caption{(a) Behavior of $r_{\mathrm{rms}}(z)$ of the symmetric generated beams with varying $\delta_a=\delta_b=\delta_{a,b}$ for $\Omega=\Gamma$ and $\Gamma/4$ (inset). (b) Longitudinal parameters for the symmetric generated beams for the same values of $\delta_{a,b}$. Red and black curves correspond to $\Omega=\Gamma$ and $\Omega=\Gamma/4$, respectively.}
\label{fig:gaussian1}
\end{figure}

\begin{figure}[t!]
\includegraphics[width = 1\linewidth]{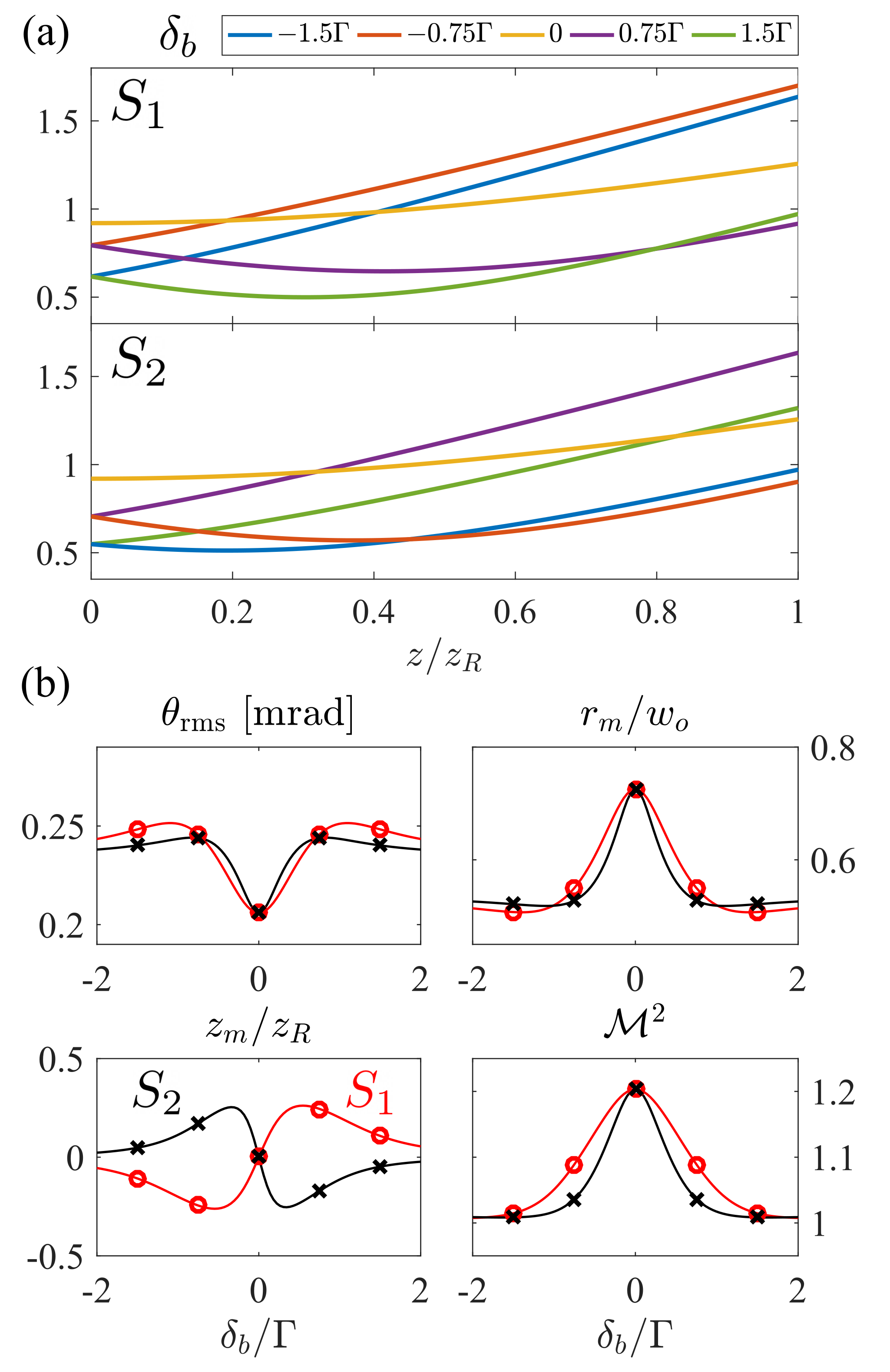}
\centering
\caption{(a) Behavior of $r_{\mathrm{rms}}(z)$ of signals $S_1$ (top) and $S_2$ (bottom), and (b) longitudinal parameters of $S_1$ (red) and $S_2$ (black) for $\delta_a = 0$, and different values of $\delta_b/\Gamma$. Rabi frequencies are $\Omega_{a,b}^0=\Gamma$.}
\label{fig:dda_fixed_res}
\end{figure}

To better understand the influence of the medium on the spatial characteristics of the generated beams, we investigate the effect of the detunings from resonance, $\delta_{a,b}$, on the free space propagation of the generated beam.
First, we consider a situation where both beams have equal detunings, i.e., come from the same laser source.
We see from Fig. \ref{fig:gaussian1} that $\delta_{a,b}$ has an intuitive effect on the FWM beams.
On resonance, the beams are generated with the maximum radius and the minimum divergence.
As we move away from the resonance, the radius right outside the sample decreases while the divergence angle increases.
It is interesting to note that $z_m$ changes considerably for varying incident beam detunings.
Above resonance, $z_m$ is shifted to negative values, while below resonance, it is shifted to positive values.
These results suggest that it is possible to translate the position where the minimum waist of the FWM beam occurs by controlling the frequency of the incident beams.
This translation comes with not much change on the other beam parameters.
Another parameter is the beam quality factor, which is maximum at resonance and approaches unity as $\delta_{a,b}$ goes away from resonance.

To evidence that the influence of the frequency degrees of freedom depends on the incident beam amplitudes, we show in the inset of Fig. \ref{fig:gaussian1}(a) the rms radii of the symmetric generated beams when $\Omega_{a,b}^0=\Omega=\Gamma/4$.
We see that all five curves, at each $\delta_{a,b}$ value, are now closer to each other.
The longitudinal parameters for this low Rabi frequency are shown in the black curves of Fig. \ref{fig:gaussian1}(b), and we can see that $r_m$ varies much less with the detunings and $z_m$ stays around $0$ for all $\delta_{a,b}$.
%In the limit $\Omega_{a,b}^0\rightarrow0$, all five curves in Fig. \ref{fig:gaussian1}(a) would become equal, and so would the values of the longitudinal parameters at all detunings $\delta_{a,b}$.
This is because, in the limit $\Omega^0_{a,b}\rightarrow0$, the coupling $\kappa$ becomes uniform and factors out of the integral in Eq. (\ref{eq:Lambda}). As a result, the frequency degrees of freedom would not affect the overall shape of the generated beam, only the power conversion efficiency.
Moreover, we can see that the beam quality factor satisfies $\mathcal{M}^2\geq1$ for all detunings and approaches the value $\mathcal{M}^2=1$ for increasing $|\delta_{a,b}|$.

We now turn to a situation where the two incident beams do not have the same detuning but maintain the Rabi frequencies equal, $\Omega_a^0 =\Omega_b^0$.
In this case, the symmetry under the exchange $a\leftrightarrow b$ in Eq. (\ref{eq:s12}) no longer holds, and the generated signals $S_1$ and $S_2$ are shown to differ.

First, we set the frequency of $E_a$ on resonance, $\delta_a=0$, and make $\delta_b$ vary around $\delta_b=0$.
Figure \ref{fig:dda_fixed_res}(a) shows the radii of both generated beams on free-space propagation in this situation.
We see that by varying only the detuning $\delta_b$, we also obtain changes in the focusing region of both FWM beams.
However, the positions of minimum radius, $z_m$, of the two signals are translated to opposite directions (Figure \ref{fig:dda_fixed_res}(b)).
A more focused beam on direction $(2\mathbf{k}_a-\mathbf{k}_b)$ is accompanied by a more spread beam on direction $(2\mathbf{k}_b-\mathbf{k}_a)$, and vice versa.
The values of the divergence angle $\theta_{\mathrm{rms}}$, minimum radius $r_m$, and quality factor $\mathcal{M}^2$, also shown in Figure \ref{fig:dda_fixed_res}(b), are very similar for both signals.
Further, their dependence on $\delta_b$ is similar to that seen in the case with $\delta_a=\delta_b$.

\begin{figure}[t!]
    \centering
    \includegraphics[width=1\linewidth,trim={0cm 0.2cm 0cm 0cm}]{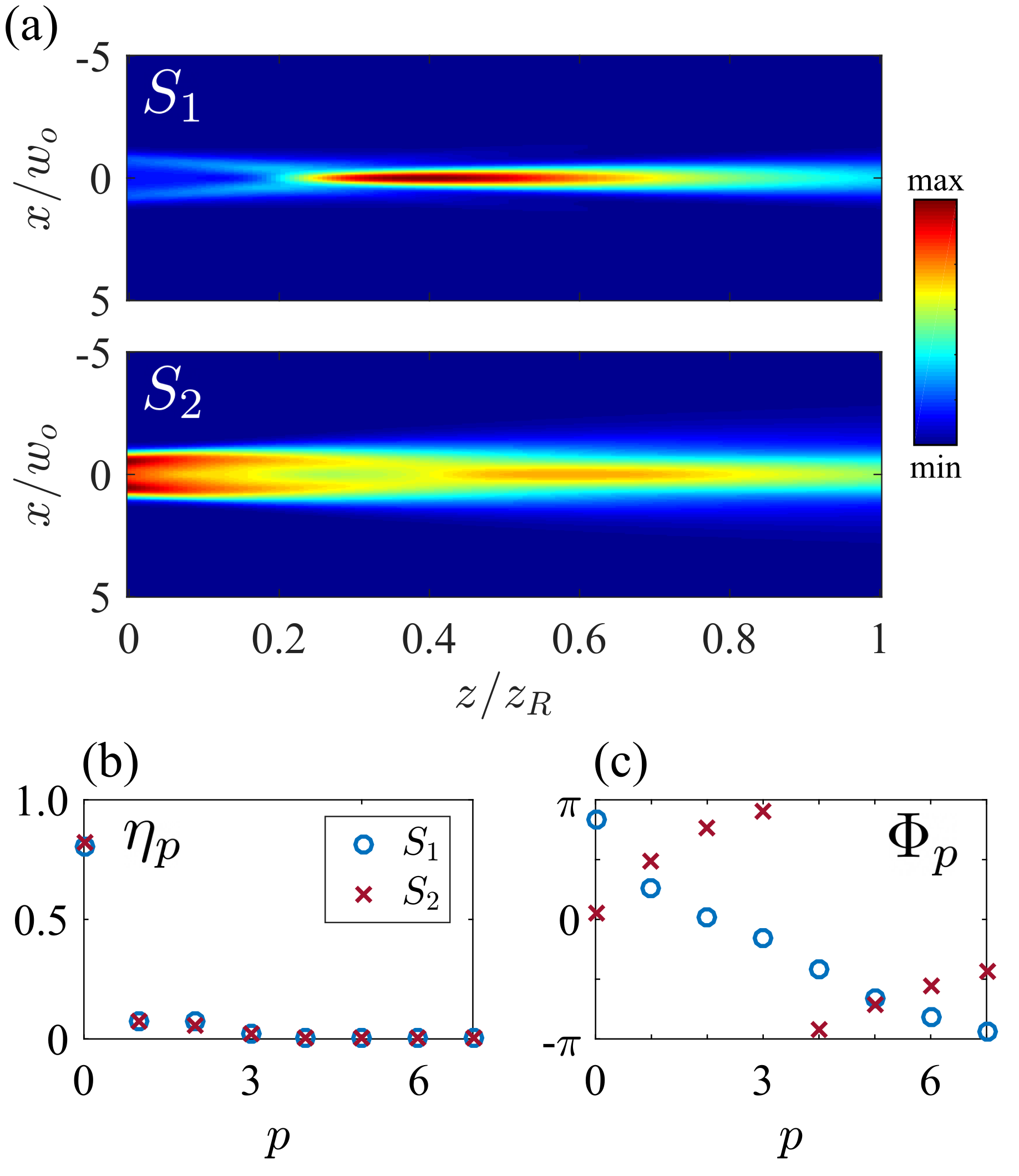}
    \caption{(a) Propagation of both generated beams outside the nonlinear medium from $z/z_R=L/2z_R\approx0$ to $z/z_R=1$, (b) mode components ${\eta}_{p}$ and (c) phases $\Phi_{p}$ of $S_1$ and $S_2$ for $\delta_a=0$ and $\delta_b=0.75\Gamma$, corresponding to the purple curves of Fig. \ref{fig:dda_fixed_res}(a).}
    \label{fig:dda_fixed_res_prop}
\end{figure}

In Figure \ref{fig:dda_fixed_res_prop}(a) we show the longitudinal profile of $S_1$ and $S_2$ for $\delta_a=0$ and $\delta_b=0.75\Gamma$, corresponding to the purple curves of Figure \ref{fig:dda_fixed_res}(a).
Referring to the discussion regarding the position of maximum intensity, we see that in this case, the position of minimum $r_{\mathrm{rms}}(z)$ seems to be closer to the position of maximum intensity outside the nonlinear medium.
The distributions of $\eta_{p}$, shown in Fig. \ref{fig:dda_fixed_res_prop}(b) are similar for both signals, with $u_{0}$ having the greatest contribution, as before, and slightly different weights for the modes with $p\neq0$.
However, the distribution of relative phases of the modes $u_{p}$ that are being superimposed, shown in Fig. \ref{fig:dda_fixed_res_prop}(c), is quite different between $S_1$ and $S_2$.
This is the dominant factor that leads to the differences in the longitudinal profiles of Fig. \ref{fig:dda_fixed_res_prop}(a).

Our calculations reveal an effect that resembles Kerr lensing \cite{NLOboyd}, where the total index of refraction in the medium can be written as $n=n_0+n_2I$, where $n_0$ and $n_2$ are the linear and nonlinear refractive indices and $I$ is the beam intensity.
The total index of refraction seen by the strong beam is thus modulated by its own spatial distribution.
Evidently, the generated beam is not strong enough to induce self-modulation.
However, we can explain qualitatively the observed focusing/defocusing effect by expressing the total index of refraction seen by the generated field as
\begin{equation}
n_t = n_0 + \Delta n(I_a,I_b;\delta),    
\end{equation}
where $\Delta n(I_a,I_b;\delta)\propto\mathrm{Re}\chi^{2a-b}(r;\delta)$ is a nonlinear contribution modulated by the spatial distribution of both incident beams.

\begin{figure}[h!]
    \centering
    \includegraphics[width=1\linewidth]{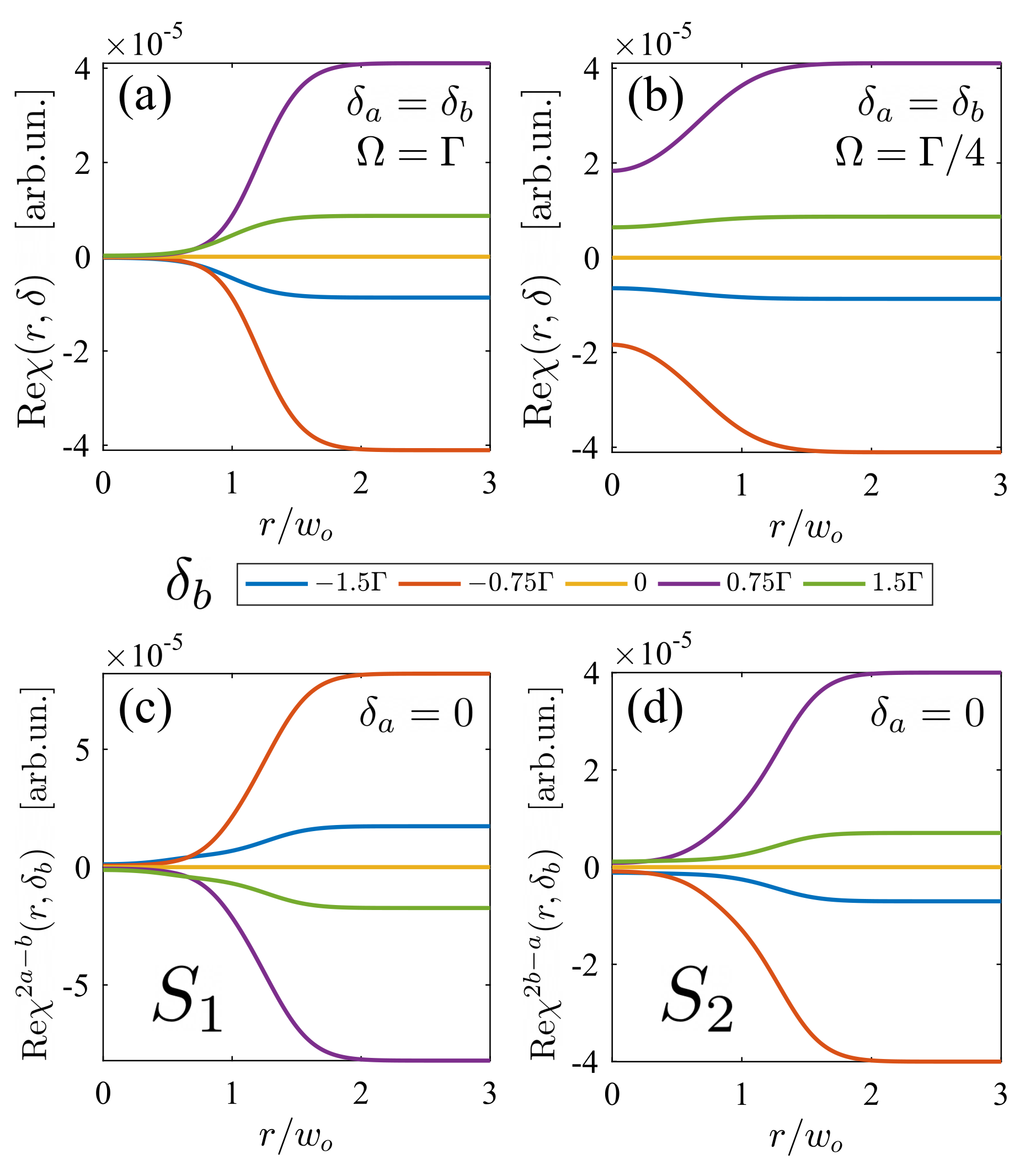}
    \caption{Radial distributions of the real part of the nonlinear susceptibility for different cases. For (a)-(b) the detunings are equal $\delta_a=\delta_b=\delta$, depicting the completely symmetric case, $\chi^{2a-b}=\chi^{2b-a}=\chi$. Rabi frequencies are (a) $\Omega^0_{a,b}=\Gamma$ and (b) $\Omega^0_{a,b}=\Gamma/4$. For (c)-(d) $\delta_a$ is fixed on resonance, and we have the non-symmetric case. In (c) and (d) we show respectively $\mathrm{Re}\chi^{2a-b}(r;\delta_b)$ and $\mathrm{Re}\chi^{2b-a}(r;\delta_b)$ with $\Omega^0_{a,b}=\Gamma$.}
    \label{fig:radial_chi3}
\end{figure}

Now we look at the first situation considered: incident fields with equal detunings, $\delta_a=\delta_b$, and symmetric nonlinear signals.
We show in figures \ref{fig:radial_chi3}(a) and \ref{fig:radial_chi3}(b) plots of the real part of the nonlinear susceptibility as a function of the radial coordinate for different values of $\delta_{a}=\delta_b$.
The Rabi frequencies are $\Omega_{a,b}^0=\Gamma$ (Fig. \ref{fig:radial_chi3}(a)) and $\Omega_{a,b}^0=\Gamma/4$ (Fig. \ref{fig:radial_chi3}(b)), corresponding to the situations of Fig. \ref{fig:gaussian1}(a) and its inset, respectively.
We see that below resonance, $\delta_{a,b}<0$ (red and blue curves), the total index of refraction is greater at the center, $r=0$, and decreases at greater $r$ positions.
Thus, the FWM beam is focused.
On resonance, $\delta_{a,b} = 0$, the total index of refraction is unaffected by $\chi^{2a-b}=\chi^{2b-a}=\chi$ at all radial positions.
The FWM beam is neither focused nor defocused.
Above resonance, $\delta_{a,b}>0$ (purple and green curves), $n_t$ is smaller at the center, and increases as we move away from this position.
As a result, the FWM beam is defocused.
For the lower intensity (Fig. \ref{fig:radial_chi3}(b)) the variation of the refractive index is smaller. We then expect the (de)focusing effect to be weaker.
This agrees with the behavior of $z_m$ shown in Fig. \ref{fig:gaussian1} and its inset.

The same analysis can be made regarding the second case, where $\delta_a$ is fixed at resonance, and $\delta_b$ varies. As already discussed, the two FWM signals in this case are non-symmetric. In figures \ref{fig:radial_chi3}(c) and \ref{fig:radial_chi3}(d) we present plots of the real part of the nonlinear susceptibility associated with the generation of FWM signals $S_1$ ($2\mathbf{k}_a-\mathbf{k}_b$) and $S_2$ ($2\mathbf{k}_b-\mathbf{k}_a$) with the input Rabi frequency $\Omega^0_{a,b}=\Gamma$.
The nonlinear susceptibility related to $S_2$ has a radial dependence similar to that shown in Fig. \ref{fig:radial_chi3}(a)-(b) for $\delta_b>0$ and $\delta_b<0$, while for $S_1$, the curves are switched.
This indicates that the (de)focusing effect is opposite between $S_1$ and $S_2$, in agreement with the results for $z_m$ shown in Fig. \ref{fig:dda_fixed_res}(b).
The opposite behavior of the two generated beams is related to whether the beams that participate with one and two photons in the nonlinear process have negative or positive detunings.

When the incident fields carry topological charge, the radial profile of the susceptibility is further affected by their non-Gaussian intensity profiles.
More interestingly, the phase distribution of the generated field is twisted when the detunings are varied around resonance \cite{hamedi2018exchange,yu2021engineering}.
The phase discontinuities that arise from the azimuthal phase factors $e^{\pm i \ell \phi}$, inherent to vortex beams, are the features that reveal this twisting effect.
We are conducting a separate study that deals with this situation.

\section{Conclusions} \label{conclusions}

We have shown that the combined spatial and spectral degrees of freedom of the nonlinear susceptibility in a thin medium lead to intuitive effects on the free space propagation properties of the generated FWM beam.
In particular, we analyzed a situation where two nonlinear signals are induced by the same incident beams, and studied their transverse and longitudinal characteristics.
The FWM fields were calculated as superpositions of paraxial modes, with the expansion coefficients given by the overlap integral of the nonlinear polarization.
Effects of incident beam power and detunings from resonance were investigated.

For increasing power, we showed that the distribution of coefficient amplitudes is sensibly affected, leading to greater contributions from higher radial order modes.
The intensity profile at the medium exit was shown to suffer notable change, with an intensity ring being formed as the pump power increases.
Physically this can be understood as a spatial saturation effect.
This intensity ring is not stable, and under free propagation to distances of the order of the Rayleigh range outside the interaction medium, the ring-shaped profile transitions into one that is Gaussian at the center.

When the detunings are varied around resonance, the distribution of phases of the superimposed modes is mainly affected.
In this case, the FWM beam is shown to undergo focusing or defocusing, depending on whether the fields are above or below resonance.
We argued that it is possible to identify a nonlinear contribution to the refractive index inside the atomic medium, similar to the Kerr effect.
The fundamental difference is that this nonlinear index is affected by the intensity distribution of the strong pump beams, and not by the generated beam itself.
Moreover, in the configuration where two FWM signals are generated simultaneously, we showed that the effect can be either symmetric between the two beams, if the detunings are equal, or anti-symmetric, if the detunings are different, e.g., one of the detunings fixed on resonance.

The transfer of transverse structure of light beams in nonlinear processes is strongly dictated by the overlap of incident beams, and this assumption leads to remarkably accurate theoretical predictions both in third- and second-order processes.
However, by taking into account the spatial structure of the medium susceptibility, effects due to pump power and frequency arise, and we show that in this case the overall spatial shape of the generated signal can be sensibly affected.
The results presented in this work indicate that the FWM beam structure, in the near-field, can reveal characteristics of the nonlinear light-atom interaction, such as the resonances of the medium, saturation effects and nonlinear variations of the refractive index; while after propagation to distances of the order of a Rayleigh range, this information fades away as the generated beam evolves into a stable form, which is strongly dictated by the overlap of incident beam modes.

\appendix

\section{Steady state OBEs} \label{appendix:RHO}

The modeling of the atomic quantities is based on that of Ref. \cite{boyd81}, where the usual case of the nonlinear process involving a strong pump and a weak probe is considered.
In our specific setup both fields possess similar intensities, and thus we look for solutions to the relevant matrix elements of the density operator that are influenced by both pumps.
The set of steady state equations for the Fourier components in Eqs. (\ref{eq:coh_2lvl}) and (\ref{eq:pd_2lvl}) is

\begin{align} \label{eq:obe_2lvl}
    (\Delta\rho)^{a-b}&=\dfrac{2i\widetilde{\Omega}_{a}^{*}\sigma_{12}^{2a-b}+2i\widetilde{\Omega}_{b}^{*}\sigma_{12}^{a}-2i\widetilde{\Omega}_{a}\sigma_{12}^{b*}}{\left(i\delta_{a}-i\delta_{b}+\Gamma\right)},
    \\
    (\Delta\rho)^{\mathrm{dc}}&=(\Delta\rho)^{0}-\dfrac{4}{\Gamma}(\mathrm{Im}[\Omega_{a}^{*}\sigma_{12}^{a}]+\mathrm{Im}[\Omega_{b}^{*}\sigma_{12}^{b}]),
    \\
    \sigma_{12}^{a}&=\dfrac{i\widetilde{\Omega}_{a}(\Delta\rho)^{\mathrm{dc}}+i\widetilde{\Omega}_{b}(\Delta\rho)^{a-b}}{\left(i\delta_{a}+\Gamma/2\right)},
    \\
    \sigma_{12}^{b}&=\dfrac{i\widetilde{\Omega}_{b}(\Delta\rho)^{\mathrm{dc}}+i\widetilde{\Omega}_{a}\left[(\Delta\rho)^{a-b}\right]^*}{\left(i\delta_{b}+\Gamma/2\right)},
    \\
    \sigma_{12}^{2a-b}&=\dfrac{i\widetilde{\Omega}_{a}(\Delta\rho)^{a-b}}{\left(2i\delta_{a}-i\delta_{b}+\Gamma/2\right)}.
\end{align}
With a direct substitution method, it is possible to arrive at the slow coherence $\sigma_{12}^{2a-b}$ shown in the main text. The coherence related to signal $S_2$, in direction $(2\mathbf{k}_b - \mathbf{k}_a)$, can be found by introducing in Eq. (\ref{eq:coh_2lvl}) a term $\rho_{12}^{2b-a}$ that oscillates at $(2\omega_b - \omega_a)$ and following the same procedure. It is evident that this will result in a solution to the slow coherence $\sigma^{2b-a}_{12}$ that has the same form as $\sigma_{12}^{2a-b}$ with the exchange of labels $a\leftrightarrow b$.

\section{Calculation of the rms parameters} \label{appendix:rms}

Following Ref. \cite{vallone16} we can write the rms radius of field $\mathcal{E}_s=\sum_p\mathcal{A}_p u_p$ as
\begin{equation*} \label{eq:Rrms}
\begin{split}
    r_{\mathrm{rms}}(z)&=\frac{w(z)}{\sqrt{2}}\left[1+\braket{N}-\mathrm{Re}\left\{\varphi e^{2i\tan^{-1}(z/z_R)}\right\}\right]^{\frac{1}{2}},
\end{split}
\end{equation*}
where $\braket{N}=\sum_{p}{\eta}_{p}2p$ is the mean value of the mode order $N_{p}=2p$ in the superposition and $\varphi = \frac{1}{P}\sum_{p>0} 2p\mathcal{A}_{p}\mathcal{A}_{p-1}^*$ is a generally complex-valued factor.
By straightforward manipulations we can obtain the form shown in Eq. (\ref{eq:rms_par}) of the main text, $r_\mathrm{rms}(z)=\sqrt{r^2_m+\theta^2_\mathrm{rms}(z-z_m)^2}$, where we have explicitly
\begin{align}
    r^2_m &= \frac{w^2_o}{2}\frac{[1 + \braket{N}]^2 - |\varphi|^2}{1+\braket{N}+\mathrm{Re}\{\varphi\}},
    \\
    z_m &= -z_R\frac{\mathrm{Im}\{\varphi\}}{1+\braket{N}+\mathrm{Re}\{\varphi\}},
    \\
    \theta_{\mathrm{rms}} &= \dfrac{w_o}{\sqrt{2}z_R}\left[1+\braket{N}+\mathrm{Re}\{\varphi\}\right]^{\frac{1}{2}}.
\end{align}

\begin{acknowledgments}
This work was supported by CAPES (PROEX 534/2018, No.
23038.003382/2018-39). The authors would like to acknowledge the financial support from Brazilian agencies CAPES and CNPq. M. R. L. da Motta acknowledges financial support from CNPq (130306/2020-7) and CAPES (88887.623521/2021-00).  A. A. C. de Almeida acknowledges financial support by CNPq (141103/2019-1).
\end{acknowledgments}

%\nocite{*}

\bibliography{apssamp}% Produces the bibliography via BibTeX.

\end{document}